\documentclass[SIL, reviewmode]{AEA}

\usepackage{natbib}

\draftSpacing{1.5}

\begin{document}

\title{Time-consistent decisions and rational expectation equilibrium existence in DSGE models}
\shortTitle{Time consistency and equilibrium existence}
\author{Minseong Kim}
\date{\today}
\pubMonth{}
\pubYear{}
\pubVolume{}
\pubIssue{}

\begin{abstract}
 Under some initial conditions, it is shown that time consistency requirements prevent rational expectation equilibrium (REE) existence for dynamic stochastic general equilibrium models induced by consumer heterogeneity, in contrast to static models. However, one can consider REE-prohibiting initial conditions as limits of other initial conditions. The REE existence issue then is overcome by using a limit of economies. This shows that significant care must be taken of when dealing with rational expectation equilibria.
\end{abstract}

\maketitle

\section{Introduction}
In a dynamic stochastic general equilibrium (DSGE) model, satisfaction of time consistency \citep{kydland77} is crucial, especially for a rational expectation equilibrium (REE).

Without stochastic shocks, DSGE models are equivalent to multi-period static models, except for the question of time consistency, which prohibits some of possible equilibria in dual multi-period static models.

This paper intends to demonstrate that under some initial conditions, there is no time-consistent REE in an unexpected way. There is surprise in that conventional dynamic model analysis \citep{slp89} suggests there should be no equilibrium existence issue. The resolution to this `paradox' is easily found: for an economy $E_0$, if we instead take limit $E \rightarrow E_0$, then there exists the limit of model-corresponding equilibria. There is no problem utilizing this limit equilibrium instead. This limit strategy evades the time consistency issue by taking the limit of economies that do not have time consistency issues.

\section{Time consistency issue}
In fashion of a New Keynesian model \citep{woodford03,gali15}, assume that a central bank controls issuance of $B_{it}$, where $i$ refers to each consumer. When $B_{it}>0$, consumer has government bonds, when $B_{it} <0$, consumer is indebted to the central bank. $B_{it}$ is the only way consumers can build up wealth or be indebted. As standard with many DSGE models, an infinite-horizon economy is assumed.

Suppose that the central bank, for whatever reasons, decides to impose $B_{it} = 0$, along with a right interest rate path for the implementation. That is, central bank decides to run an economy as if there is no finance. While the assumption is quite toxic, this is not expected to produce no equilibrium by itself.

Now the specification of consumer $i$ follows.
\begin{equation}
\label{eqn:girutilitymax}
\max_{\{C_{it}, B_{it},..\}}U_i = \max_{\{C_{it}, B_{it},..\}} \sum_{t=0}^{\infty} (\beta_i)^t u_i(C_{it},..)
\end{equation}
where $C_{it}$ refers to consumption good, $\beta_i$ refers to time discount factor and $u_i$ refers to a single-period utility function with $U$ being time-discounted utility function. It is assumed that generally, $\beta_i \neq \beta_j$. Let the budget constraint of each consumer be:
\begin{equation}
\label{eqn:girbudget}
P_t C_{it} + (1+i_t)^{-1} B_{it} +.. \leq  B_{i(t-1)}+..
\end{equation}
where $..$ terms do not contain any of $C_{it}$ and $B_{it}$. $P_t$ refers to price level and $i_t$ is nominal interest rate set by central bank.

The following main equation may be derived from consumer optimization, with the derivation provided in the appendix:
\begin{equation}
\label{eqn:girmain}
\frac{u_j'(C_{jt})}{u_k'(C_{kt})} = \frac{(1-\gamma_{jk})}{\gamma_{jk}}\left(\frac{\beta_k}{\beta_j}\right)^t
\end{equation}
Note that Equation \eqref{eqn:girmain} can be derived even when $B_{it}$ and central bank do not exist - see the appendix.

Suppose that economic conditions remain time-invariant. For example, there is zero technology factor growth, there is no capital accumulation in an economy as standard in a basic New Keynesian model \citep{woodford03,gali15} and so on. As aforementioned, $B_{it} = 0$, and there are infinitely many time periods. Then Equation \eqref{eqn:girmain} creates the time inconsistency issue not found in a dual multi-period static model.

Since economic conditions remain same across periods and there is no `end' in time, the left-hand side of Equation \eqref{eqn:girmain} should remain constant across time. This is impossible when $\beta_k \neq \beta_j$.

\subsection{Limit economy}
Note that the `initial condition' requirement for the time inconsistency result is relatively strong. First, $B_{it} = 0$ was required. Second, other economic conditions also are assumed to remain same across periods. An infinite horizon also plays a role - however, technically, the time inconsistency result itself can be created even in a finite horizon circumstance.

If any of these conditions is not satisfied, then Equation \eqref{eqn:girmain} cannot be asserted to generate the time consistency issue.

We may then choose to approach an economy with the time consistency issue as a limit of neighboring economies that do have time-consistent equilibria. A limit of equilibria then can be asserted to be an equilibrium for a limit economy (the economy with the time consistency issue) as well.

Therefore, in practical sense, an economy with the time consistency issue may safely be treated as if it has an equilibrium. In case of the example economy, even if fiscal authority does not actually exercise its power, its potential power to reallocate $B_{it}$ helps to entirely avoid theoretical issues. 

\section{Conclusion}
The lesson is that there must be significant care that must be taken of when dealing with rational expectation equilibria (REE). It is well-known that not all REE are meaningful or learnable. \citep{evans01,evans18} The direction also goes from non-existence as well - non-existence of REE has to be carefully considered.

For REE non-existence driven by a time consistency issue induced from consumer heterogeneity, it is easy to construct nearby economies that do not have time consistency issues, thereby evading the issue. In such a case, it is not economics-wise meaningful to declare that there exists no REE.

\bibliographystyle{aea}
\bibliography{nkrec}

\appendix
\section{Derivation of the main equation}
Set up a Lagrangian:
\begin{equation}
\label{eqn:girlagrangian}
U_j + \sum_{t=0}^{\infty}\left[\lambda_{jt} \left(B_{j(t-1)} + .. - \left[P_t C_{jt} + (1+i_t)^{-1}B_{jt} + ..\right]\right)\right]
\end{equation}
Let us consider the first-order condition associated with $B_{jt}$. It goes:
\begin{equation}
\label{eqn:girmid1}
\lambda_{j(t+1)} - \lambda_{jt} (1+i_t)^{-1} = 0
\end{equation}
thus
\begin{equation}
\label{eqn:connect}
\frac{\lambda_{j(t+1)}}{\lambda_{jt}} = \frac{1}{1+i_t}
\end{equation}
Consider the first-order condition associated with $C_{jt}$. It goes:
\begin{equation}
\label{eqn:girmid3}
(\beta_j)^t u_j'(C_{jt}) - \lambda_{jt}P_t = 0
\end{equation}
thus
\begin{equation}
\label{eqn:girmid4}
\lambda_{jt} = \frac{(\beta_j)^t u_j'(C_{jt})}{P_t}
\end{equation}
Now pick two consumers $j$ and $k$:
\begin{equation}
\label{eqn:girmid5}
\frac{\lambda_{jt}}{\lambda_{kt}} = \frac{(\beta_j)^t u_j'(C_{jt})}{(\beta_k)^t u_k'(C_{kt})}
\end{equation}
\begin{equation}
\label{eqn:girmid6}
\frac{u_j'(C_{jt})}{u_k'(C_{kt})} = \frac{(\beta_k)^t}{(\beta_j)^t}\frac{\lambda_{jt}}{\lambda_{kt}}
\end{equation}
Because of Equation \eqref{eqn:connect},
\begin{equation}
\label{eqn:girmid7}
\frac{\lambda_{jt}}{\lambda_{kt}} = \frac{\lambda_{j(t+1)}}{\lambda_{k(t+1)}} = \frac{1-\gamma_{jk}}{\gamma_{jk}}
\end{equation}
for some constant $\gamma_{jk}$. Therefore, we obtain Equation \eqref{eqn:girmain}.

\section{Other ways the main equation may be derived}
Because the derivation of the main equation relied on existence of a central bank bond $B_{it}$, we would prefer the derivation that reproduces Equation \eqref{eqn:girmain} without having to rely on existence of $B_{it}$. This can be done by re-adapting the idea in \cite{negishi60}.

Negishi (1960) states that a candidate competitive equilibrium has to be a solution of a social planner problem with the utility function that assigns constant weights to the utility function of individual agent.

But here, we are looking at general DSGE models, so there are cases when firms are not price-taking and et cetera. So can we generalize Negishi (1960)? The answer is yes.

The proof in Negishi (1960) only requires that ``feasible'' utility vectors form a convex set and equilibria to sit on the boundary of the set. 

By a utility vector, it means $\bar{U} = (U_1,U_2,..)$, where subscript indices refer to consumers/agents.

In original Negishi (1960), ``feasible'' is defined as resource-wise feasible. But this does not need to be the case. If we can redefine the word ``feasible'' - such as considering how firms behave to set price that would prevent additional possible allocations - we may allow feasible utility vectors to form a convex set, while equilibria sit on the boundary of the set.

In such a case, let $U_s$ be:
\begin{equation}
\label{eqn:negishiutility}
U_s = \gamma U_1 + (1-\gamma)U_2 + \sum_{j=3}^{n} \nu_j U_3
\end{equation}
where $\gamma$ and $\nu_j$ are constants, and $n$ refers to the number of consumers. Let some of sequential constraints for the obtained social planner problem at each time $t$ be of form:
\begin{equation}
\label{eqn:negishiresource}
C_{1t} + C_{2t} +... \leq ...
\end{equation}
where $...$ terms do not contain any of $C_{1t}$ and $C_{2t}$, and rest of constraints do not contain any of $C_{1t}$ and $C_{2t}$.

Then we get Equation \eqref{eqn:girmain} as the result of first-order conditions. First-order conditions say:
$$\gamma U_1'(C_{1t}) - \lambda_t = 0$$
$$(1-\gamma) U_2'(C_{2t}) - \lambda_t = 0$$
where $\lambda_t$ is the sum of Lagrange multipliers attached to constraints that share the form of Equation \eqref{eqn:negishiresource}. We can rewrite the above equations as:
$$\gamma (\beta_1)^t u_1'(C_{1t}) = \lambda_t$$
$$(1-\gamma) (\beta_2)^t u_2'(C_{2t}) = \lambda_t$$
Thus:
$$\gamma (\beta_1)^t u_1'(C_{1t}) = (1-\gamma) (\beta_2)^t u_2'(C_{2t})$$
which gives us Equation \eqref{eqn:girmain}.

While the required form of constraints to derive Equation \eqref{eqn:girmain} seem stringent, it actually is not. One way to see this is as follows. From an initially competitive competition economy, we may change firms to be monopolistically competitive. But in such cases, consumer profiles do not change. Unless firms engage in price discrimination against consumers, derived constraints will follow the form of Equation \eqref{eqn:negishiresource} or will not have $C_{1t}$ and $C_{2t}$ terms.

\end{document}